\def\be{\begin{equation}}
\def\ee{\end{equation}}
\def\bea{\begin{eqnarray}}
\def\eea{\end{eqnarray}}
\def\a{\alpha}
\def\b{\beta}

\def\L{\Lambda}
\def\p{\psi}
\def\r{\rho}
\def\t{\theta}

\def\s{\sigma}
\def\o{\omega}
\def\hr{{\widehat R}}
\def\hd{{\widehat D}}
\def\ha{{\widehat A}}
\def\hk{{\widehat K}}
\def\hi{{1}}
\def\hb{{\widehat B}}
\def\hu{{\widehat U}}
\def\hv{{\widehat V}}
\def\hw{{\widehat W}}
\def\hht{{\widehat T}}

\documentclass[12pt]{article}

\usepackage{amsmath, amsgen,amstext,amsbsy,amsopn,amsthm, amssymb }
\begin{document}

\title{The one-dimensional Hubbard model:\\
 A reminiscence}

\author{E. H. Lieb$^1$ and F. Y. Wu$^2$ \\
$^1$Departments of Mathematics and Physics, Princeton University\\
Princeton, New Jersey, 08544 USA \\
$^2$Department of Physics, Northeastern University \\ Boston, 
Massachusetts 02115, USA}
 
\renewcommand{\thefootnote}{}
\footnotetext{\copyright\, 2002 by the authors. This paper may be 
reproduced, in its entirety, for non-commercial purposes.}

\date{July 21, 2002}
\maketitle
\begin{abstract}

In 1968 we published the solution of the ground state energy and wave
function of the one-dimensional Hubbard model, and we  also showed
that there is no Mott transition in this model.  Details of the
analysis have never been published, however.  As the Hubbard model has 
become increasingly important in condensed matter physics, relating
to topics such as the theory of high-$T_c$ superconductivity, it is
appropriate to revisit the one-dimensional model  and to recall here
some details of  the solution.

\end{abstract}

\vskip 1cm
\newpage
\section{Introduction}
In a previous paper \cite{liebwu} we reported the solution of the
one-dimensional (1D) Hubbard model, showing the absence of the Mott
transition in its ground state, but the letter format of the paper did
not permit the presentation of all the details of the analysis.  Over
the years the Hubbard model \cite{gutzwiller, hubbard} has become more
important, for it plays an essential role in several topics in
condensed matter physics, including 1D conductors and high-$T_c$
superconductivity. It also plays a role in the chemistry of aromatic
compounds (e.g., Benzene \cite{heil,nach}). Several books
\cite{gaud,ha,mont,taka} now exist in which the 1D Hubbard model is
analyzed, and numerous papers have been written on properties of the
model \cite{phyrev}.  Almost invariably these publications are based
upon results of \cite{liebwu}, including the absence of a Mott
transition, but without derivation. While other rigorous results on
higher dimensional Hubbard models exist, and some of these are
reviewed in \cite{rev,tas}, the ID model stands as the only Hubbard
model whose ground state can be found exactly.  It has been brought to
our attention that it would be useful to students and researchers if
some details of the solution could be made available.  Here, taking
the opportunity of the symposium, StatPhys-Taiwan 2002,
which takes place in the year when
both of the authors turn 70, we revisit the 1D Hubbard model and
present some details of the 34-year old solution.

While our paper \cite{liebwu} contained significant  results about the
excitation spectrum, it was mainly concerned with the integral
equations for the ground state and we concentrate on those equations
here. The new, unpublished results are contained in Sects. 5, 6, 7.

Consider a crystal of $N_a$ lattice sites with a total of $N$
itinerant electrons hopping  
between the Wannier states of neighboring lattice sites, and that each
site is capable of accommodating two electrons of opposite spin, with
an interaction energy $U>0$, which mimics a screened Coulomb repulsion
among electrons.  The Hubbard model \cite{hubbard} is described by the
Hamiltonian
\begin{eqnarray}
\label{Hubbard}
{\cal H} \, = \, \, T \, \sum_{<ij>} \sum_{\sigma} \,
 c_{i\; \sigma}^{\dag}\; c_{j\; \sigma}\; \, + \, U \, \sum_{i} \;
n_{i\; \uparrow }\; n_{i\; \downarrow}\; , 
\end{eqnarray}
where $\, c_{i\; \sigma}^{\dag}\;$ and $\, c_{i\; \sigma}\, $ are,
respectively, the creation and annihilation operators for an electron
of spin $\, \sigma\, $ in the Wannier state at the $\, i$-th lattice
site and $n_{i\; \s} = c_{i\; \s}^{\dag}\; c_{i\; \s}\;$ is the
occupation number operator.  The summation $<ij>$ is over nearest
neighbors, and one often considers (as we do here) periodic boundary
conditions, which means that $<ij>$ includes a term coupling opposite
edges of the lattice.  We are interested in the ground state solution
of the Schr\"odinger equation ${\cal H}\; \big| \p\rangle = E \;\big|
\p\rangle \, $.

For bipartite lattices (i.e., lattices in which the set of sites can
be divided into two subsets, $A$ and $B$, such that there is no
hopping between $A$ sites or between $B$ sites), such as the
one-dimensional chain, the unitary transformation $V^\dagger\, {\cal
H}\, V $ leaves ${\cal H}$ unchanged except for the replacement of $T$
by $-T$.  Here $V={\rm exp} \,\big[i\pi \sum_{i\in A}(n_{i\; \uparrow
}+n_{i\; \downarrow })\big]$, with $A$ being one of the two
sublattices.  Without loss of generality we can, therefore, take
$T=-1$. In any event, bipartite or not, we can renormalize $U$ by
redefining $U$ to be $U/|T|$.  Henceforth, the value of $T$ in
(\ref{Hubbard}) is $-1$ and $U$ is positive and fixed.  The dependence
of the Hamiltonian and the energy on $U$ will not be noted explicitly.

The commutation relations
\begin{equation*}
\Big[\sum_i n_{i\; \uparrow },\ {\cal H}\Big] =
\Big[\sum_i n_{i\; \downarrow }, \ {\cal H}\Big] =0 
\end{equation*}
imply that the numbers of down-spin electrons $M$ and up-spin
electrons $M'$ are good quantum numbers.  Therefore we characterize
the eigenstates by $M$ and $M'$, and write the Schr\"odinger equation
as
\begin{equation}
\label{Sequation}
{\cal H} \;\big|  M, M'\rangle =E (M, M')\; \big| M, M'\rangle\, .
\end{equation}
Naturally, for any fixed choice of $M, M'$ there will generally be
many solutions to (\ref{Sequation}), so that $\big| M, M'\rangle $ and
$E(M,M')$ denote only generic eigenvectors and eigenvalues.
Furthermore, by considering particles as holes, and vice versa,
namely, introducing fermion operators
\begin{equation*}
d_{i\; \s} = c_{i\; \s}^\dag\, , \hskip 1cm
d_{i\; \s}^\dag = c_{i\; \s}\,  ,
\end{equation*}
and the relation $n_{i\; \s} = 1- d_{i\; \s}^\dag\, d_{i\; \s}$,
we obtain the identity
\begin{equation}
\label{energyinvert}
 E(M, M')= -(N_a-N)U+ E (N_a-M,N_a- M')
\end{equation}
where 
\begin{equation*}
N=M+M' 
\end{equation*}
 is the total number of electrons.
Since  $N \geq N_a$ if, and only if, $(M-N_a)+(M'-N_a) \leq N_a$, 
we can   restrict our considerations to
\begin{equation*}
N\leq N_a, 
\end{equation*}
namely, the case of at most a ``half-filled band''.  In addition,
owing to the spin-up and spin-down symmetry, we need only consider
\begin{equation*}
M \leq M'. 
\end{equation*}

\section{The one-dimensional model}\label{onedimen}
We now consider the one-dimensional model, and write
 $\big| M, M'\rangle$ 
as a linear combination
of states with  electrons  at specific sites. 
Number the lattice sites by $1, 2, \cdots, N_a$ and, since we
want to use periodic boundary conditions,  we require $N_a$ to be an even
integer in order to retain the bipartite structure. For later use it is
convenient also to require that $N_a = 2 \times ({\mathrm{odd\ integer}})$
in order to be able to have $M=M'=N_a/2$ with $M$ odd.  For the 1D
model the sum in (\ref{Hubbard}) over $<ij>$ is really a sum over $1\leq i
\leq N_a, j=i+ 1$, plus $1\leq j \leq N_a, i=j+ 1$ with $N_a+1 \equiv 1$.

Let $ \big|\; x_1,  
  \,\cdots\, ,  x_N\rangle$ denote the state in which 
the   down-spin electrons are located at  sites
$\, x_1, \, x_2, \,
\cdots\,, x_M\, $ and the up-spin electrons are  at sites $\, x_{M+1},
 \,\cdots\, ,  x_N$.
  \medskip
The eigenstate is now written as
\begin{equation}
\label{eigenstate}
\big|\, M, M'\rangle = \sum_{1\leq x_i \leq N_a} f(x_1,
  \,\cdots\, ,  x_N)\big|\; x_1,  
  \,\cdots\, ,  x_N\rangle
\end{equation}
where the summation is over all $\,x_1, \cdots, x_N\,$ from
$1$ to $N_a$, and $\, f(x_1, 
  \,\cdots\, ,  x_N) \, $ is the amplitude of the state $\big|\; x_1,  
  \,\cdots\, ,  x_N \rangle$.

It is convenient to denote the $N$-tuple $x_1, x_2, ..., x_N$ simply
by $X$.

By substituting (\ref{eigenstate}) into the Schr\"odinger equation
(\ref{Sequation}), we obtain (recall $T=-1$)
\begin{eqnarray}
\label{Gaudin}
 &&-\sum_{i=1}^{N}\Big[ f(\,x_1,   \cdots , x_{i} 
 +1 ,  \cdots ,  x_N) +  f(\,x_1,   \cdots , x_{i} 
 -1 ,  \cdots\, ,  x_N)\Big] \nonumber  \\
&&\quad + \, U \,\Big[ \sum_{i<j} 
\delta(x_{i}\, -x_{j}) \Big]\,  f(\,x_1,  \,\cdots ,  x_N) \, 
= \,\, \, E\, f(\,x_1, \,\cdots\, ,  x_N),
\end{eqnarray}
where $\delta $ is the Kronecker delta function.  We must solve
(\ref{Gaudin}) for $f$ and $E$, with the understanding that site $0$
is the same as site $N_a$ and site $N_a+1$ is the same as site $1$
(the periodic boundary condition). Equation (\ref{Gaudin}) is the
`first quantized' version of the Schr\"odinger equation
(\ref{Sequation}). It must be satisfied for all $1\leq x_i\leq N_a$,
with $1\leq i \leq N$.

As electrons are governed by Fermi-Dirac statistics, we require that
$\, f(X)$ be antisymmetric in its first $\, M$ and last $\, M'$
variables {\it separately}. This antisymmetry also ensures that $f=0$
if any two $x$'s in the same set are equal, which implies that the
only delta-function term in (\ref{Gaudin}) that are relevant are the
ones with $i\leq M$ and $j > M$. This is consistent with the
definition of ${\mathcal H}$ in (\ref{Hubbard}), in which the only
interaction is between up- and down-spin electrons.

The antisymmetry allows us 
to reinterpret (\ref{Gaudin}) in the following alternative way.
Define the region $R$ to be the following 
subset of all possible values of $X$ (note the $<$ signs): 
\bea
\label{regime}
 R\;= \quad \left\{X\; :\;
\left(
\begin{array}{c}
 \quad 1\leq \, x_1< x_2 < \cdots < x_M \leq N_a \\
 1 \leq   x_{M+1} < x_{M+2} < \cdots < x_N \leq N_a 
\end{array}
\right)\;\right\}\, .
\eea
In $R$ any of the first $M$ $x_i$'s  can be equal to any of the
last $M'$, with an interaction energy $nU$, where $n$ is the number of
overlaps of the first set with the second.

The antisymmetry of $f$ tells us that $f$ is completely determined by
its values in the subset $R$, together with the requirement that $f=0$
if any two $x$'s in the same set are equal (e.g.,
$x_1=x_2$). Therefore, it suffices to satisfy the Schr\"odinger
equation (\ref{Gaudin}) when $X$ on the right side of (\ref{Gaudin}) is
only in $R$, together with the additional fact that we set $f=0$ on the
left side of (\ref{Gaudin}) if $x_i\pm 1$ takes us out of $R$, e.g.,
if $x_1+1 =x_2$. ({\it Warning:} With this interpretation, equation
(\ref{Gaudin}) then becomes a self-contained equation in $R$ alone and
one should not ask it to be valid if $X\notin R$.)

There is one annoying point about restricting attention to $R$ in
(\ref{Gaudin}). When $x_1=1$ the left side of (\ref{Gaudin}) asks for
the value of $f$ for $x_1=N_a$, which takes us outside $R$. Using the
antisymmetry we conclude that
\begin{equation}\label{pbc}
f(N_a, x_2, ...,x_N)= (-1)^{M-1} f(x_2, ...,x_{M}, N_a, 
x_{M+1},...,x_N),
\end{equation}
with similar relations holding for $x_M=1$, $x_{M+1}=N_a$ \, or
$x_N =1$. Equation (\ref{pbc}) and its three
analogues  reflect the ``periodic boundary conditions''
and, with its use, (\ref{Gaudin}) becomes  a self-contained equation on
$R$ alone. 

We now come to the main reason for introducing $R$. {\it Let us assume
that $M$ and $M'$ are both odd integers.} Then $(-1)^{M-1} =
(-1)^{M'-1} =1$ and we claim that: {\it For all $U$, the ground state
of our Hamiltonian satisfies}

\begin{enumerate}
\item There is only one ground state and
\item $f(X)$ is a strictly positive function in $R$.
\end{enumerate}

To prove 2. we think of (\ref{Gaudin}) as an equation in $R$, as
explained before.  We note that all the off-diagonal terms in the
Hamiltonian (thought of as a matrix $\widehat{\mathcal H}$ from
$\ell^2(R)$ to $\ell^2(R)$) are non-positive (this is where we use the
fact that $(-1)^{M-1} = (-1)^{M'-1}=1$). If $E_0$ is the ground state
energy and if $f(X)$ is a ground state eigenfunction (in $R$), which
can be assumed to be real, then, by the variational principle, the
function $g(X)= |f(X)|$ (in $R$) has an energy at least as low as that
of $f$, i.e., $\langle g\,|\, g\, \rangle = \langle f\,|\, f\, \rangle
$ and $\langle g |
\widehat{\mathcal H} | g \rangle \leq \langle f | \widehat{\mathcal H}
| f \rangle $ since  $ |f(X)|\, \widehat{\mathcal H}(X,Y) \, |f(Y)|
\leq f(X) \widehat{\mathcal H}(X,Y) f(Y)$ for every $X,Y$.  Hence $g$
must be  a ground state as well (since it cannot have a lower energy
than $E_0$, by the definition of $E_0$).  Therefore, $g(X)$ must satisfy
(\ref{Gaudin}) with the same $E_0$. Moreover, we see from (\ref{Gaudin})
that $g(X)$ is strictly positive for {\it every} $X\in R$ (because if
$g(Y)=0$ for some $Y\in R$ then $g(Z)=0$ for every $Z$ that differs from
$Y$ by one `hop'; tracing this backward, $g(X)=0$ for every $X\in R$).

Returning now to $f$, let us assume  the contrary of 2., namely,  $f(X)>0$
for some $X\in R$, and $f(Y)\leq 0$ for some $Y\in R$.
We observe that since $h=g-f$ must also be a ground state
(because sums of ground states are ground states, although not necessarily
normalized), we have a ground state (namely $h$) that is non-negative
and non-zero, but not strictly positive; this contradicts the fact,
which we have just proved, that every non-negative ground state must be
strictly positive. Thus, 2. is proved.

A similar argument proves 1.  If $f$ and $f'$ are two linearly
independent ground states then the state given by $k(X)=f(X) +cf'(X)$
is also a ground state and, for suitable $c$, \ $k(X)=0$ for some
$X\in R$, but $k$ cannot be identically zero. Then $|k|$ will be a
non-negative ground state that is not strictly positive, and this
contradicts statement 2.

The uniqueness statement 1. is important for the following reason.
Suppose that we know the ground state for some particular value of $U$
(e.g., $U=\infty$) and suppose we have a $U$-dependent solution to
(\ref{Gaudin}) in some interval of $U$ values (e.g., $(0,\infty)$)
with an energy $E(U)$ such that: (a) $E(\infty)$ is the known ground state
energy and (b) $E(U)$ is continuous on the interval. Then $E(U)$ is
necessarily the ground state energy in that interval. If not, the
curve $E(U)$ would have to cross the ground state curve (which is
always continuous), at which point there would be a degeneracy --
which is impossible according to 1.

Items 1. and 2. can be used in two main applications.
The first is the
proof of the fact that when $M$ and $M'$ are odd the ground state
belongs to total spin $S$ equal to $|M - M'|/2$ and not to some higher $S$
value. The proof is the same as in \cite{liebmattis}. In
\cite{liebmattis} this property was shown to hold for {\it all} values
of $M$ and $M'$, but for an open chain instead of a closed chain. In
the thermodynamic limit this distinction is not important.

The second main application of these items 1. and 2. is a proof
that the state we construct below using the Bethe Ansatz really is the
ground state. This possibility is addressed at the end of Sect.
\ref{Bethesec} where we outline a strategy for such a proof. 
Unfortunately, we are unable to carry it out and we leave it as an
open problem.

We also mention a theorem \cite{lieb2}, which states that {\it the
ground state is unique for $M = M' = N_a/2$ and $N_a =$ even (the
half-filled band).}  There is no requirement for  $M=M'$ to be odd.

\section{The  Bethe Ansatz}\label{Bethesec}
The Bethe Ansatz was invented \cite{bethe} to solve the Heisenberg
spin model, which is essentially a model of lattice bosons. The boson
gas in the continuum  with a positive delta function interaction and
with positive density in the thermodynamic limit was first treated in
\cite{lieb}. McGuire \cite{mcguire} was the first to 
realize that the method could be extended to continuum fermions with a
delta function interaction for $M =1$.  (The case $M =0$ is trivial.) 
The first real mathematical difficulty comes with $M =2$ and this was
finally solved in
\cite{flicker}. The solution was inelegant and not transparent, but
was a precursor to the full solution for general $M$ by Gaudin
\cite{gaudin} and Yang \cite{yang}.

We now forget about the region $R$ and focus, instead, on the
fundamental regions (note the $\leq$ signs)
\begin{equation}
\label{Q}
R_Q = \left\{ X\; :\; 1\leq x_{Q1} \leq \cdots \leq x_{QN} \leq N_a
\right\}
\end{equation}
Here $Q=\{Q1, Q2, \cdots, QN \}$ is the permutation that maps the
ordered set $\{1,2, ...,N\}$ into $\{Q1, Q2, ..., QN\}$. There are
$N!$ permutations and corresponding regions $R_Q$. The union
of these regions is the full configuration space. These regions
are disjoint {\it except} for their boundaries (i.e., points where
$x_{Qi}= x_{Q(i+1)}$).

Let $k_1<k_2<...<k_N$  be a set of {\it unequal, ordered} 
and {\it real} numbers
in the interval $-\pi <k \leq \pi$, and
let $[Q,P]$ be a set of $N! \times N!$ coefficients indexed by a pair
of permutations $Q,P$, all yet to be determined.

When $X\in R_Q$ we write the function $f(X)$ as (the Bethe Ansatz)
\begin{equation}
\label{ansatz}
f(X)=f_Q(\,x_1,  \,\cdots ,  x_N) = \sum_P \,[Q,P]\, {\rm exp} \big[ i
\big(k_{P1}x_{Q1}
+ \cdots + k_{PN}x_{QN} \big)\big]\, .
\end{equation}

In order for (\ref{ansatz}) to represent a function on the whole
configuration space it is essential that the definitions (\ref{ansatz})
agree on the intersections of different $R_Q$'s. This will impose
conditions
on the $[Q,P]$'s.

Choose some integer $1\leq i<  N$ and let $j=i+1$. Let $P, P'$ be two
permutations such that $Pi=P'j$ and $Pj = P'i$, but otherwise $Pm =
P'm$ for $m\neq i,j$. Similarly, let $Q,Q'$ be a pair with the same
property (for this same choice of $i$) but otherwise $P,P'$ and $Q,Q'$
are unrelated.

The common boundary between $R_Q$ and $R_{Q'}$ is the set in which
$x_{Qi} = x_{Qj}$. In order to have $f_Q = f_{Q'}$ on this boundary 
it is sufficient
to require that
\begin{equation}
\label{cont}
[Q,P] + [Q,P'] = [Q',P] + [Q',P']  \ .
\end{equation}
The reason that  this suffices is that on this boundary we have
$x_{Qi}=x_{Qj}$ and $k_{Pi} +k_{Pj} = k_{P'i} +k_{P'j}$. 
Thus, (\ref{cont}) expresses the fact that
 the exponential factor $\exp[i(k_{Pi}x_{Qi} + k_{Pj}x_{Qj})] =
\exp[i(k_{P'i}x_{Qi} + k_{P'j}x_{Qj})]$ is the same for
$Q$ and $Q'$, and for all values of the other $x_m$'s.

Next we substitute the Ansatz (\ref{ansatz})
into (\ref{Gaudin}). 
If   $\big|x_i-x_j\big|>1$  for all $i,j$ then, 
clearly, we have
\begin{equation}
\label{energy}
E=E(M, M')= -2 \sum_{j=1}^N \cos \,k_j.
\end{equation}
We next choose the coefficients $[Q,P]$ to make (\ref{energy})  hold
generally --- even if it is not possible to have $\big|x_i-x_j\big|>1$
for all $i,j$ when the number of electrons exceeds $N_a/2$. The requirement
that (\ref{energy}) holds will impose further conditions on $[Q,P]$ similar to 
(\ref{cont}).

Sufficient conditions are obtained by setting $x_{Qi} = x_{Qj}$ on the
right side of (\ref{Gaudin}) and requiring the exponential factors
with $x_{Qi}$ and $x_{Qj}$ alone to satisfy (\ref{Gaudin}). In other
words, we require that
\begin{eqnarray}\label{encond} \nonumber
& &[Q,P]\,e^{-ik_{Pj}}+[Q,P']\,e^{-ik_{P'j}}+[Q,P]\,e^{+ik_{Pi}}+
[Q,P']\,e^{+ik_{P'i}} \\ \nonumber
&=& [Q',P]\,e^{-ik_{Pi}}+[Q',P']\,e^{-ik_{P'i}}+[Q',P]\,e^{+ik_{Pj}}+
[Q',P']\,e^{+ik_{P'j}} \\
& & \hskip 1cm +U\Big([Q,P]+[Q,P']\Big) \ .
\end{eqnarray}
If we combine (\ref{encond}) with (\ref{cont}) and recall that 
$ k_{Pj}= k_{P'i}$, etc., we obtain
\begin{eqnarray}
\label{yangeq}
[Q,P]&=& \frac{-iU/2}{\sin k_{Pi}  - \sin k_{Pj} +iU/2}\, [Q,P']  \nonumber \\
   && \hskip  2cm +\,\frac{\sin k_{Pi} - \sin k_{Pj}}{\sin k_{Pi} - \sin k_{Pj} +iU/2}
 \,[Q',P'] \ .
\end{eqnarray}

It would seem that we have to solve both (\ref{yangeq}) {\it and}
(\ref{cont}) for the $(N!)^2$ coefficients $[Q,P]$, and for each
$1\leq i\leq N-1$. Nevertheless, (\ref{yangeq}) alone is sufficient
because it implies (\ref{cont}). To see this, add (\ref{yangeq}), as given,
to (\ref{yangeq}) with $[Q',P]$ on the left
side. Since $Q'' = Q$, the result is (\ref{cont}). Our goal, then, is
to solve ({\ref{yangeq}) for the coefficients $[Q,P]$ such that the
amplitude $f$ has the required symmetry.

These equations have been solved in \cite{gaudin} and \cite{yang},
as we stated before, and we shall not repeat the derivation here. In
these papers the function $\sin k$ appearing in (\ref{yangeq}) is
replaced by $k$, which reflects the fact that 
\cite{gaudin} and \cite{yang} deal with
the continuum and we are working on a lattice. This makes no difference
as far as the algebra leading to equations (\ref{equations}) is
concerned, but it makes a big difference for constructing a proof that
these equations have a solution (the reason being that the sine
function is not one-to-one). 

The algebraic analysis in \cite{gaudin} and \cite{yang}
leads to the following set of $N+M$ equations
for the $N$ ordered, real, unequal $k$'s. (Recall that
$M\leq M'$.) They involve an additional set of $M$ ordered, unequal
real numbers $\L_1 < \L_2 < \cdots < \L_M$.
\bea\label{equations}
&& e^{i k_j N_a} = \prod_{\b=1}^M \frac {i\sin \, k_j -i \L_\b -U/4}
{i\sin \, k_j -i \L_\b +U/4}\, , \quad j=1,2,\cdots, N \label{betheeq}  \\
&&\prod _{j=1}^N \frac {i\sin \, k_j -i \L_\a -U/4} {i\sin \, k_j -i 
\L_\a +U/4}\,
=-\, \prod_{\b=1}^M \frac {-i\L_\b + i \L_\a +U/2}{-i\L_\b + 
i \L_\a -U/2}\, , 
\quad \a = 1,2,\cdots,M. \nonumber 
\eea
We remark that an explicit expression for the wave function $f(X)$ has
been given by Woynarovich \cite[part 1, Eqs. 2.5-2.9]{woyna}.

These equations can be cast in a more transparent form (in which we now
really make use of the fact that the $k$'s and $\L$'s are ordered) by
defining
\begin{equation*}
\t(p) = -2 \tan^{-1} \Big( \frac{2p} {U} \Big), \hskip 1cm -\pi \leq 
\t \leq \pi 
\end{equation*}
Then, taking the logarithm of (\ref{betheeq}), we obtain  two sets of 
equations
\begin{eqnarray}
 && \hskip 1cm N_ak_j = 2 \pi I_j +\sum_{\b=1}^M \t\big(2\sin\, k_j - 2\L_\b\big), \quad 
j=1,2,\cdots, N \label{finiteequ1}  \\ 
&&\sum_{j=1}^N \t\big(2\sin\, k_j - 2\L_\a\big) = 2\pi J_\a 
-\sum_{\b=1}^M\t\big(\L_\a-\L_\b\big),\;
\a = 1,2,\cdots, M \quad\quad \label{finiteequ2}
\end{eqnarray}
where $I_j$ is an integer (half-odd integer) if
$M$ is even (odd), while  $J_\a$  is an integer (half-odd integer)
if $M'$ is odd (even). 

It is noteworthy that in the $U\to\infty $ limit the two sets of
equations essentially decouple. The $\L$'s are proportional to $U$ in
this limit, but the sum in (\ref{finiteequ1}) becomes independent of
$j$. In particular, when the $\L$'s are balanced (i.e., for every $\L$ there
is a $-\L$) as in our case, then  this sum equals zero. 

{}From (\ref{finiteequ1}) and (\ref{finiteequ2}) we have   the identity
\begin{equation}\label{ktotal}
k_{\rm total}\equiv\sum_{j=1}^N k_j =
 \frac {2 \pi} {N_a} \Big( \sum_{j=1}^N I_j + \sum_{\a=1}^M J_\a\Big)
\end{equation}
For the ground state, with $N= 2\times$ (odd integer)
and $M=N/2=$ odd, we make
the choice of the $I_j$ and $J_\alpha$ that agrees with the correct
values in the case $U= \infty$, namely
\begin{equation}
I_j = j-(N+1)/2   \quad\quad\quad J_\alpha = \alpha -(M+1)/2
\end{equation}

We are  not able to prove the existence of solutions to
(\ref{finiteequ1}) and (\ref{finiteequ2}) that are real and
increasing in the index $j$ and $\alpha$. In the next section,
however, we show that the $N\to \infty$ limit of
(\ref{finiteequ1}) and (\ref{finiteequ2}) has a solution, and in
Sect. \ref{half} we obtain the solution explicitly for $N/N_a
=1$. This leaves little doubt that (\ref{finiteequ1})
and(\ref{finiteequ2}) can be solved as well, at least for large $N$.

Assuming that $M= M'=N/2$ is odd, the solution is presumably unique
with the given values of $I_j$ and $J_\alpha$ and belongs to
total spin $S=0$.

Assuming that the solution exists, we would still need a few more
facts (which we have not proved)
in order to prove that the Bethe Ansatz gives the ground state.

(a.) Prove that the wave function (\ref{ansatz}) is not identically zero.

(b.) Prove that the wave function (\ref{ansatz}) is a continuous
function of $U$.

\noindent
{}From the uniqueness of the ground state proved in
Sect. \ref{onedimen}, and the fact that the solution (\ref{ansatz})
coincides with the exact solution for $U=\infty$ (in which case $f_Q(x)$
is a Slater determinant of plane waves with wavenumbers
$k_j=2\pi I_j/N_a$), (a.) and (b.) now establish that the wave function
(\ref{ansatz}) must be the ground state for all $U$. 

{\it Remark:\/} Assuming that the Bethe Ansatz gives the ground state
for a given $M \leq M'$ then, as remarked at the end of Sect. 2 (and
assuming $M$ and $M'$ to be odd) the value of the total spin in this
state is $S= (M'-M)/2$. Thus, the solution to the Bethe Ansatz we have
been looking at is a highest weight state of $SU(2)$, i.e., a state 
annihilated by spin
raising operators.

\section{The ground state}
For the ground state $I_j=I(k_j)$ and $J_\a=J(\L_\a)$ are consecutive
integers or half-odd integers centered around the origin.
As stated in Sect. \ref{Bethesec},
each $k_j$ lies in $[-\pi,\pi]$ (since $k_j\to k_j +2\pi n$ defines the
same wave function). In the limit of $N_a,\, N,\, M,\, M' \to \infty$ with
their ratios kept fixed, the real numbers $k$ and $\L$ are distributed
between $-Q$ and $Q\leq \pi $ and $-B$ and $B\leq \infty$ for some $0<
Q\leq \pi$ and $0 < B \leq \infty$. In a small interval $dk$ the number
of $k$ values, and hence the number of $j$ values in (\ref{finiteequ1}),
is $N_a \rho(k)dk$, where $\rho$ is a density function to be determined.
Likewise, in a small interval $d\L$ the number of $\L$ values and $\alpha$
values in (\ref{finiteequ2}) is $N_a \sigma(\L) d\L$.  An alternative
point of view is to think of $I(k)$ as a function of the
variable $k$. Then $I(k+dk)-I(k)$ counts the number of $k$ values between
$k$ and $k+dk$ so we have ${\rm d}I(k)/{\rm d}k = N_a \rho(k)$. A similar
remark holds for $J(\L)$.

The density functions $\r(k)$ and $\s(\L)$ satisfy the obvious
normalization
\begin{equation}
\label{density}
\int_{-Q}^Q \r(k) dk = N/N_a\, ,\hskip 1cm \int_{-B}^B \s(\L) d\L =
M/N_a.
\end{equation}
By subtracting (\ref{finiteequ1}) with $j$ from (\ref{finiteequ1})
with $j+N_a \rho(k)dk$, and taking the limit $N_a\to \infty$ we obtain
(\ref{contequ1}) below. Likewise, subtracting
(\ref{finiteequ2}) with $\alpha$ from (\ref{finiteequ2}) with
$\alpha+N_a \sigma(\L)d\L$, and taking the limit $N_a\to \infty$ we
obtain (\ref{contequ2}). An alternative point of view is to take the
derivatives of (\ref{finiteequ1}) and (\ref{finiteequ2}) with respect
to $k_j$ and $\Lambda_\alpha$ respectively, set
$dI/dk = N_a\rho(k)$, $dJ/d\L = N_a \s(\L)$,  and take the $N_a \to
\infty$ limit.
\bea
&& \quad \quad 1 = 2\pi \, \r(k) + 2\cos k \int_{-B}^B d\L \,\s(\L)
 \, \t'\big(2\sin\, k- 2\L\big)\; , \label{contequ1} \\
&& -2 \int_{-Q}^Q dk \,\r(k)  \, \t'\big(2\sin\, k- 2\L\big) \nonumber \\
 && \hskip 2cm = 2 \pi\, \s(\L) -  \int_{-B}^B d\L'\, \s(\L')  
 \, \t'(\L-\L') \label{contequ2}
\eea
or, equivalently, 
\begin{equation}\label{rho} 
\boxed{\quad \r(k) =
 \frac 1{2\pi} + \cos k \int_{-B}^BK(\sin k - \L\,) \s(\L)\, d\L\; ,\quad }
\end{equation}
\begin{equation}\label{sigma}
\boxed{\quad \s(\L) =\int_{-Q}^QK(\sin k-\L)\, \r(k)\, dk 
    - \int_{-B}^B K^2(\L- \L')\, \s(\L')\,  d\L' \, ,\quad }
\end{equation}
where
\bea
K(\L- \L') &=&  - \frac 1{\pi}\, \t'(2\L-2\L') = 
        \frac 1{2\pi} \Big[ \frac {8U}{U^2+16(\L-\L')^2}\Bigg]  \nonumber \\
 K^2(\L- \L') &=& - \frac 1{2\pi}\, \t'(\L-\L')
=\frac 1 {2\pi}\Bigg[ \frac {4U}{U^2+4(\L-\L')^2}\Big] \nonumber  \\
 &=&\int_{-\infty}^\infty K(\L-x)K(x-\L')\, dx \, . \nonumber
\eea
Note that $K^2$ is the square of $K$ in the sense of operator
products. Note also that  (\ref{rho}) and (\ref{sigma}) are
to be satisfied only for $|k| \leq Q$ and $|\L| \leq B$. Outside these
intervals $\rho$ and $\s$ are not uniquely defined, but we
{\it can} and will define them by the right sides of  (\ref{rho}) and (\ref{sigma}).

The following Fourier transforms will be used in later discussions: 
\begin{equation}
\label{FT}
\int_{-\infty}^{\infty} e^{i\omega \L} K(\L)\,d\L = e^{-U|\omega|/4},\quad
\int_{-\infty}^{\infty} e^{i\omega \L} K^2(\L)\,d\L = e^{-U|\omega|/2}\, .
\end{equation}

The ground state energy (\ref{energy}) now reads
\begin{equation}
\label{energy1}
E(M,M') = -2 N_a \int_{-Q}^Q \r(k) \cos k \, dk.
\end{equation}
where
$\r(k)$ is to be determined together with $\s(\L)$ from the coupled
integral equations (\ref{rho}) and (\ref{sigma}) 
subject to the normalizations (\ref{density}).

\section{Analysis of the integral equations}\label{analysis}
In this section we shall prove that equations (\ref{rho}) and
(\ref{sigma}) have unique solutions for each given $0< Q\leq \pi$ and
$0<B\leq \infty$ and that the solutions are positive and have certain
monotonicity properties. These properties guarantee that the
normalization conditions (\ref{density}) uniquely determine values of
$Q,B$ for each given value of $N$ when $M=M'=N/2$ (in this case
we have $B=\infty$). However, we have not proved uniqueness of $Q,B$ when
$M \neq M'$ (although we believe there is uniqueness).   But this does
not matter for the absolute ground state since, as remarked earlier, the
ground state has $S=0$ (in the thermodynamic limit) and so we are allowed to
take $S^z=0$.
For $M \neq M'$, we have remarked  earlier that the solution 
 probably has $S=|M'-M|/2$ and  is the ground state
 for $S=|M'-M|/2$.
 
An important first step is to overcome the annoying fact (which is
relevant for $Q>\pi/2$) that $\sin k$ is not a monotonic function of
$k$ in $[-\pi, \pi]$. To do this we note that $(\cos k) K(\sin k -
\L)$ is an odd function of $k-\pi/2$ (for each $\L$) and hence $\r(k)
-1/2\pi$ also has this property. On the other hand, $K(\sin k, \L)$
appearing in (\ref{sigma}) is an even function of $k-\pi/2$. As a
result $\r(k)$ appearing in the first term on the right side of
(\ref{sigma}) can be replaced by $1/2\pi$ in the intervals $Q'<k < Q$
and $- Q<k < -Q'$, where $Q'=\pi-Q$. Thus, when $Q>\pi/2$, we can
rewrite the $[Q',Q]$ portion of the first integral in (\ref{sigma}) as
\begin{equation*}
\int_{Q'}^Q  K(\sin k -\L) \r(k)\, dk =  
\int_{Q'}^{Q} K(\sin k -\L) \frac{1}{2\pi} \, dk
= \frac{2}{2\pi} \int_{Q'}^{\pi/2} K(\sin k -\L) \, dk
\end{equation*}
A similar thing can be done for the $[-Q,-Q']$ portion and for the 
corresponding portions of 
(\ref{density}). 

The integrals over $k$ now extend at most  over the interval 
$[-\pi/2, \pi/2]$, in which $\sin k$ is monotonic.

We are now in a position to change variables as follows. For 
$-1 \leq x\leq 1$ let
\begin{equation}\label{frho}
t(x) = \frac1{2\pi}(1-x^2)^{-1/2}, \quad f(x) = 
(1-x^2)^{-1/2}\rho (\sin^{-1}x )\ .
\end{equation}
In case $Q<\pi/2$, $\r(\sin^{-1}x)$ is defined only for $\sin x\leq
Q$, but we shall soon see (after (\ref{feqn}) 
how to extend the definition of $f$ in this
case. We define the step functions for all real $x$ by
\begin{align} \label{step}
B(x) &= 1,   \ &|x|&<B\, ;  \hskip 1.5truecm &= 0,\quad &otherwise \nonumber \\ 
A(x)  &= 1, \  &|x|&<a\, ;  \hskip 1.5truecm &= 0,\quad &otherwise \nonumber \\ 
D(x)  &= H(Q),\  &a<|x|&<1\, ; \hskip 1.5truecm &= 0,\quad &otherwise
\end{align}
where $a=\sin Q=\sin Q'$ and where
\begin{equation*}
H(Q) = 0 \quad {\rm {if}}\  Q\leq \pi/2, \qquad\qquad 
H(Q) = 2 \quad {\rm {if}}\ Q>\pi/2\ .
\end{equation*}
The integral equations (\ref{rho}) and (\ref{sigma})  become
\bea
f(x) &=& t(x) +\int_{-\infty}^{\infty}K(x-x')B(x') \s(x')dx', 
\hskip 2truecm |x|\leq a \label{feqn}\\
\s(x) &=& \int_{-\infty}^{\infty}K(x-x')A(x') f(x') dx' 
+ \int_{-\infty}^{\infty}K(x-x')D(x') t(x') dx' \nonumber \\
&& \quad -\int_{-\infty}^{\infty}K^2(x-x')B(x')\s(x') dx' 
\hskip 2.5truecm |x|<B  \label{sigeqn}\ . 
\eea
Although these equations have to be solved in the stated intervals we
can use their right sides to define their left sides for {\it all}
real $x$. We define $t(x)\equiv 0$ for $|x|>1$. 
It is obvious that the extended equations have (unique) solutions
if and only if the original ones do. Henceforth, we shall understand
the functions $f$ and $\s$ to be defined for all real $x$.

These equations read, in operator form, 
\bea
f &=& t + {\hk}{\hb}\,\s \label{operatorf} \\
\s &=&{ {\hk}{\ha}}\, f +{\hk} {\hd} \,t -{\hk}^2{\hb}\,
 \s \label{operators}
\eea
where ${\hk}$ is convolution with $K$ and $\ha,\hb,\hd$ are the
multiplication operators corresponding to $A,B,D$ (and which are also
projections since $\ha^2 =\ha$, etc.).

In view of the normalization requirements (\ref{density}), the space
of functions to be considered is, obviously, $L^1([-a,a])$ for $f$ and
$L^1([-B,B])$ for $\s$.  ($L^p$ is the $p-{\rm th}$ power integrable
functions and $L^\infty$ is the bounded functions.) Since $K(x)$ is in
$L^1({\mathbb R}) \cap L^\infty({\mathbb R})$, it is a simple
consequence of Young's inequality that the four integrals in
(\ref{feqn}) and (\ref{sigeqn}) are automatically in $L^1({\mathbb R})
\cap L^\infty({\mathbb R})$ when $f\in L^1([-a,a])$ and $\s\in
L^1([-B,B])$. Particular, the integrals are in $L^2({\mathbb R})$,
which allows us to define the operators in (\ref{operatorf}),
(\ref{operators}) as bounded operators on $L^2({\mathbb
R})$. 
In addition, $t$ is in $L^1({\mathbb R})$, but not in
$L^2({\mathbb R})$. To repeat, solutions in which $f$ and $\s$ are in
$L^1({\mathbb R})$ automatically satisfy $f-t$ and $\s$ are in
$L^p({\mathbb R})$ for all $1\leq p\leq \infty$.

\medskip\noindent
{\bf THEOREM 1 (Uniqueness).} {\it The solutions $f(x)$ and $\s(x)$ are unique and 
positive for all real $x$.}

\medskip\noindent
{\it REMARK:} The uniqueness implies that $f$ and $\s$ are even
functions of $x$ (because the pair $f(-x), \s(-x)$ is also a
solution). The theorem implies (from the definition (\ref{frho})) that
$\s(\L) > 0$ for all real $\L$ and it implies that $\r (k) >0$ for all
$|k| \leq \pi/2$. It does {\it not} imply that $\r (k)$, defined by
the right side of (\ref{rho}), is  non-negative for all $|k| >
\pi/2$. We shall prove this positivity, however, in Lemma 3. Note that
the positivity of $\r$ is equivalent to the statement that $f(x) <
2t(x)$ for all $|x| \leq 1$ because, from (\ref{rho}) and the evenness or $\s$, 
$\r(\pi - k) = (1/\pi)-\r(k)$.

\medskip\noindent
{\it Proof:} 
By substituting (\ref{operatorf}) into (\ref{operators}) and
rearranging slightly we obtain
\begin{equation}\label{S}
(\hi + \hk^2)\s = \hk(\ha+\hd)t + \hk^2(\hi - \hb)\s +\hk \ha \hk\hb \s\ .
\end{equation}
Since $\hk^2$ is positive definite, $\hi +\hk^2$ has an inverse
$1/(\hi +\hk^2)$, which we can apply to both sides of (\ref{S}). The
convolution operator 
\begin{equation}
\hr =\hk (\hi+\hk^2)^{-1}
\end{equation}
has a Fourier
transform $\frac1{2}{\rm sech}(\omega /4)$.  The inverse Fourier
transform is proportional to ${\rm sech} (2\pi x)$ (see
(\ref{fourier})), which is positive. In other words, $\hr$ is not only
a positive operator, it also has a positive integral kernel.

We can rewrite (\ref{S}) as
\begin{equation} \label{xi}
\left(\hi - \hw \right)\s = \hr(\ha+\hd)t \equiv \xi
\end{equation}
with 
\begin{equation}\label{W}
\hw = \hr\hk(\hi-\hb) +\hr \ha\hk\hb = \hr \left[\hk -
(\hi-\ha)\hk\hb \right]    \ .
\end{equation}
The middle expression shows that the integral kernel of $\hw$ is positive.

Clearly, $\xi >0$ as a function and $\xi \in L^1({\mathbb R})\cap
L^2({\mathbb R})$.  Also, $\hw$ has a positive integral kernel.  We note
that $\Vert \hr\Vert =1/2$ on $ L^2({\mathbb R})$ since $y/(1+y) \leq
1/2$ for $y\geq 0$. Also, $\Vert \hk \Vert =1 $, and $\Vert \hi -\hb
\Vert =1$, \ $\Vert \ha \Vert =1$, \ $\Vert \hb \Vert =1$.  In fact,
it is easy to check that $\Vert \hr \ha\hk\hb\Vert <1/2$. From this we
conclude that $\Vert \hw \Vert < 1$ on $L^2({\mathbb R})$ and thus
$\hi - \hw $ has an inverse (as a map from $ L^2({\mathbb R})\to
L^2({\mathbb R})$).

Therefore, we can solve (\ref{S}) by iteration:
\begin{equation}\label{series}
\s= \left(\hi + \hw +\hw^2 + \hw^3 + ... \, \right) \xi \ .
\end{equation}
This is a strongly convergent series in $L^2({\mathbb R})$ and hence
(\ref{series}) solves (\ref{S}) in $L^2({\mathbb R})$. It is the unique
solution because the homogeneous equation $(\hi - \hw ) \phi =0$ has
no solution.  Moreover, since each term is a positive function, we
conclude that $\s$ is a positive function as well.  {\bf QED}.

\medskip
{\bf Lemma 1 (Monotonicity in B).}  {\it When $B$ increases with $Q$
fixed, $\s(x)$ decreases pointwise for all $x \in {\mathbb R}$.}

\medskip
{\it Proof:} Since $\hi -\ha$ is fixed and positive, we see from the
right side of(\ref{W}) that the integral kernel of $\hw$ is monotone
decreasing in $\hb$. The lemma then follows from the representation
(\ref{series}).  {\bf QED}.

\medskip
{\bf Lemma 2 (Monotonicity in B).}  {\it When $B$ increases with $Q$
fixed, $f(x)$ increases pointwise for all $x\in {\mathbb R}$. This
implies, in particular, that $\rho(k)$ increases for all $|k| \leq
\pi/2$ and decreases for all $\pi/2 \leq |k| \leq \pi$.}

\medskip
{\it Proof:} Consider equation (\ref{S}) for the case $A=0$. Theorem 1
and Lemma 1 hold in this case, of course. We also note that their proofs
do not depend on any particular fact about the function $Dt$, other than
the fact that it is a non-negative function. From these observations we
learn that the solution to the equation
\begin{equation}\label{general}
(\hi + \hk^2\hb) S = \hk g
\end{equation}
has the property, for all $x\in \mathbb{R}$, that $S(x) \geq 0$ and that
$S(x)$ is a non-increasing function of $B$, provided only that $g(x)\geq
0$ for all $x\in \mathbb{R}$.

Another way to say this is that the integral kernel of $\hv= (\hi +
\hk^2\hb)^{-1}\hk$ is positive and is a pointwise monotone
decreasing function of $B$.

Now let us rewrite (\ref{general}) as 
\begin{equation}
\label{keyequation}
 {\hk^2}\, ({ \hi + \hk^2})^{-1} \, g = \hu g 
 + {\hk}\, ({ \hi + \hk^2})^{-1}\, \big( \hi - \hb\big) \, S
\end{equation}
with 
\begin{equation} \label{U}
\hu=\hk \hb \hv= \hk \hb  (\hi +
     \hk^2\hb)^{-1}\hk \ .
\end{equation}
The operator $\hu$ has a positive integral kernel since $\hv$, 
$\hk$, and $\hb$ have one.  As $B$ increases the second term on
the right side of (\ref{keyequation}) decreases pointwise (because
$(\hi- \hb)$ decreases as a kernel and $S$ decreases, as we have just
proved).  The left side of (\ref{keyequation}) is independent of $B$
and, therefore, the first term on the right side of
(\ref{keyequation}) must increase pointwise. Since this holds for
arbitrary positive $g$, we conclude that the integral kernel of $\hu$,
in contrast to that of $\hv$, is a pointwise {\it increasing} function
of $B$.

Having established the monotonicity property of $\hu$ let us return to
$f$, which we can write (from (\ref{operatorf})) as 
\bea
f &=& \big( \hi + \hu \hd \big) t + \hu \ha \, f \label{fiterate1} \\
  &=& \big[ \hi + \hu \ha +(\hu \ha)^2 + \cdots \big] (\hi + \hu \hd ) t\ .
\label{fiterate2}
\eea
The series  in (\ref{fiterate2}) is strongly convergent (since
$\Vert \ha \Vert =1$ and $\Vert \hu \Vert \leq 1/2$) and thus defines
the solution to (\ref{fiterate1}). Since $\hu$ is monotone in $\hb$, 
 (\ref{fiterate2}) tells us that $f$ is also pointwise monotone, as 
claimed.

Equation (\ref{frho}) tells us that $\rho(k)$ is increasing in $B$ for
$|k| \leq \pi/2$ and is decreasing in $B$ for
$\pi/2\leq |k| \leq \pi$  .   {\bf QED}.

\medskip
{\bf THEOREM 2 (Monotonicity in B).} {\it When $B$ increases with $Q$
fixed,  $N/N_a$ and $M/N$ increase. When $B=\infty$, 
we have $ 2M = N$, and when  $B<\infty$ we have  $ 2M <N$ (for all $Q$).}

\medskip {\it Proof:}
The integral for $N/N_a$, in (\ref{density}) can be written as
$\int_{-\infty}^\infty [A \rho + (1/2\pi)D ]$, and this is monotone
increasing in $B$ since $\rho$ is monotone for $|k| \leq 
\pi/2$ and $A(k) =0$ for $|k| > \pi/2$.

If we integrate (\ref{sigma}) from $\L=-\infty$ to $\infty$, and use
the fact that $\int K=1$ from (\ref{FT}), we obtain
\begin{equation} \label{mn}
\frac{N}{N_a}=\int_{-Q}^Q \r(k) \, dk =  \int_{-\infty}^\infty 
\s(\L) \, d\L 
 + \int_{-B}^B \s(\L) \, d\L\;
\end{equation}
which becomes, after making use of the normalization (\ref{density}),
\begin{equation}
\label{lemma1}
1 = 2 \, \frac {M}{N} +\frac{N_a}{ N}  \Bigg[ \int_{-\infty}^{-B} 
+  \int_{B}^\infty \Bigg] \s(\L)\, d\L\; .
\end{equation}
Now the integrals in (\ref{lemma1}) decrease as $B$ increases by Lemma
1 and converge to $0$ as $B\to \infty$, while $N/N_a$ increases, as we
have just proved. It follows that $M/N$ increases monotonically with
$B$, reaching $M/N=1/2$ at $B=\infty$.  If $B<\infty$ then $M/N<1/2$
since $\s$ is a strictly positive function.
\quad   {\bf QED}.

We turn now to the dependence of $\s, \rho$ on $Q$, with fixed
$B$. First, Lemma 3 (which was promised in the remark after Theorem 1)
is needed.

\medskip
{\bf Lemma 3 (Positivity of $\rho$).} {\it For all $B \leq \infty $, all
$Q\leq \pi$, and all  $|k|\leq \pi$, we have \  $\r(k) > 0$.}

\medskip\noindent
{\it Proof:} As mentioned in the Remark after Theorem 1, the positivity
of $\r$ is equivalent to the statement that $f(x) < 2\ t(x)$ for all
$|x| \leq 1$.
We shall prove  $f(x) < 2\ t(x)$ here.

Owing to the monotonicity in $B$ of $f$ (Lemma 2) it suffices to prove
the lemma for $B=\infty$, which we assume now.  We see from
(\ref{fiterate2}) that for any given value of $a$ the worst case is
$Q>\pi/2$, whence $H(Q) =2$ and $D >0$. We assume this also.

For the purpose of this proof (only) we denote the dependence of
$f(x)$ on $a$ by $f_a(x)$.

We first consider the case $a=0$, corresponding to $Q=\pi$. Let us
borrow some information from the next section, where we actually solve
the equations for $B=\infty, \ Q=\pi$ and discover (Lemma 5) that $f(x) < 2
t(x)$ for $|x|\leq 1$ (for $U >0$). 

We see from (\ref{fiterate1}) or (\ref{fiterate2}) that $f_a$ is
continuous in $a$ and differentiable in $a$ for $0<a<1$ (indeed, it is
real analytic). Also, since the kernel $K(x,y)=K(x-y)$ is smooth in $(x,y)$
and $t(x) $ is smooth in $x\in (-1,\, 1)$, it is easy to see that
$f_a$ is smooth, too, for $x\in (-1,\, 1)$. Equation (\ref{feqn})
defines $f_a(x)$ pointwise for all $x$ and $f_a(x)$ is jointly
continuous in $a,x$.

In detail, (\ref{fiterate1}) reads
\begin{equation}
\label{f}
f_a(x) =t(x)  + 2\Bigg[\int_{-1}^{-a}+\int_{a}^{1}\Bigg]{ U}(x,x')t(x')\, dx'
 + \int_{-a}^{a}{ U}(x,x')f_a(x')\,dx'\; .
\end{equation}
Take the derivative with respect to $a$ and set $h_a(x) =
\partial f_a(x)/\partial a$.  Observe that $\hu$ does not 
depend on $a$. We obtain
\bea
&&h_a(x) ={ U}(x,a)\big[f_a(a)-2t(a)\big] + { U}(x,-a)\big[f_a(-a)-2t(-a)\big] 
\nonumber \\
&&\hskip 2cm +\int_{-a}^a { U}(x,x')h_a(x') \, dx' \label{h}
\eea
(This equation makes sense because $f_a(x)$ is jointly continuous in $x,a$ and
$t(x)$ is continuous for $|x|<1$. Recall
that $f$ and $t$ are even functions of $x$. Note that $U$ here
is the kernel of (\ref{U}) with $B=\infty$, i.e., $\hu=\hk^2 (1+\hk^2)^{-1}$,
which is self-adjoint and positive as an operator and positive as a kernel.)

Equation (\ref{h}) can be iterated in the same manner as
(\ref{fiterate2}) (since $\Vert \hu \Vert =1/2$) 
\begin{equation}\label{iterate}
h_a(x) = [\hu+ \hu\ha\, \hu +\hu\ha\, \hu \ha\, \hu + \cdots](x,a) F(a) 
\equiv T(x,a)F(a)\  ,
\end{equation}\label{TF}
where $ [\ \cdot\ ](x,a)$ denotes the integral kernel of $\hht=[\
\cdot \ ]$, and where $F(a) = f_a(a)-2t(a)$ is a number. 
As an operator.  $\hht$ is self-adjoint and positive.

Now ${\hat U}$ has a positive kernel and thus $T(x,a ) \geq 0$, so
$h_a(x) <0$ for all $x$ if and only if $F(a) <0$. We have already
noted that $F(0) <0$.

We can integrate (\ref{iterate}) to obtain
\begin{equation}\label{integ}
f_a(x) = f_0(x) + \int_0^a h_{a'}(x)da' = f_0(x)+\int_0^a T(x,a')[f_{a'}(x)
-2t(a')]\, da'\ .
\end{equation}
If we subtract $2t(a)$ from this and set $x=a$, we obtain
\begin{equation}\label{G}
F(a) = G(a) + \int_0^a T(a,a')F(a')\, da'  \ ,
\end{equation}
where $G(a) = f_0(a) -2t(a) <0$. Another way to state (\ref{G}) is
$F=G+ \hht\ha F$.

Equation (\ref{G}) implies that $F(a)<0$ for all $a$, as desired. There are
two ways to see this. One way is to note that $\hht$ is monotone increasing
in $a$ (as an operator and as a kernel), so $\hht \leq \hu +\hu^2 +
\cdots =\hk^2 < \hi$. Therefore, (\ref{G}) can be iterated as
$F=[\hi+\hht\ha + \hht\ha\hht\ha + \cdots]G$, and this is negative.
The second way is to note that $f_a(a)$ (and hence $F(a)$) is
continuous in $a$.  Let $a^*$ be the smallest $a$ for which $F(a)=0$.
Then, from (\ref{G}), $0=F(a^*) = G(a^*) + \int_0^{a^*}
T(a^*,a')F(a')da' <0$, which is a contradiction.  

{}From $F(a)<0$ we can deduce that $f_a(x)-2t(x)<0 $ for all $|x|\leq1$.
Simply subtract $t(x)$ from both sides of (\ref{integ}). Then 
$f_a(x)-2t(x) = \left\{f_0(x)-2t(x)\right\} + (\hht\ha F)(x)$.
The first term $\left\{\ \right\}<0$ by Lemma 5, which we prove
in Sect. 6 below, and the second term is 
$<0$ (since $F<0$) \qquad {\bf QED}.

\medskip
{\bf Lemma 4 (Monotonicity in Q).} {\it 
 Consider the dependence of the solution to (\ref{operatorf}),
(\ref{operators}) on the parameter $0\leq a \leq 1$ for fixed $B\leq \infty$.
For  $Q\leq \pi/2$
(i.e., $H(Q) = D = 0$),
 both $f$ and $\s$ increase pointwise  as $a$ increases.
For  $Q>\pi/2$ (i.e., $H(Q) = 2, 
 Dt=2(1-A)t$),  both $f$ and $\s$ decrease
pointwise as $a$ increases.

If, instead of the dependence on $a$, we consider the dependence on
$0\leq Q\leq \pi$ of $\r(k)$ (which is defined by (\ref{rho}) for all
$|k|\leq \pi$) and of $\s(\L)$ (which is defined by (\ref{sigma}) for all
real $\L$), then, as $Q$ increases,
\begin{eqnarray}\label{rhoincr}
\r (k) && increases \ \, for\    0 \leq |k| < \pi/2 \ and \ 
decreases \ for\   \pi/2 \leq |k| \leq  \pi \nonumber \\
\s(\L) && increases \ for\ all\ real\ \L\ .
\end{eqnarray}
}
{\it Proof:} Concerning the  monotonicities stated in
the second part of the lemma, (\ref{rhoincr}), we note that as $Q$
goes from $0$ to $\pi/2$, $a$ {\it increases} from $0$ to $1$, but when $Q$
goes from $\pi/2$ to $0$, $a$ {\it decreases} from $1$ to $0$.  Moreover,
$H(Q)=0$ in the first case and $H(Q)=2$ in the second case. This observation
shows that the first part of the lemma implies the statement about
$\s$ in (\ref{rhoincr}). The statement about $\rho$ in (\ref{rhoincr})
also follows, if we take note of the $\cos k$ factor in (\ref{rho}).

We now turn to the first part of the lemma. The easy case is
$Q\leq \pi/2$ or $H(Q)=0$. Then (\ref{fiterate2}) does not have the $\hu
\hd t$ term and, since $\hu$ has a positive kernel and since $\ha $
has a kernel that increases with $a$, we see immediately that $f$
increases with $a$.  Likewise, from (\ref{xi}), (\ref{W}), we see that
$\hw$ and $\xi$ increase with $a$ and, from (\ref{series}), we see
that $\s$ increases.

For $Q>\pi/2$ or $H(Q)=2 $, we proceed as in the proof of Lemma 3 by defining 
$h_a(x) =
\partial f_a(x)/\partial a$ and proceeding to (\ref{iterate}) (but with
$\hu$ given by (\ref{U})). This time we know that $F(a) <0$ (by Lemma
3) and hence $h_a(x) <0$, as claimed. The monotonicity of $\s(x)$ follows
by differentiating (\ref{sigeqn}) with respect to $a$.  Then
$(\partial \s(x)/\partial a) = (\hv \ha h_a) (x) + V(x,a)F(a)$,
where $V(x,y) $ is the kernel of $\hv$, which is positive, as
noted in the proof of Lemma 2. \quad {\bf QED}

\medskip
{\bf THEOREM 3 (Monotonicity in Q).} {\it When $Q$ increases with
fixed $B$, \ $N/N_a$ and $M/N_a$ increase.  When $Q=\pi$, $N/N_a =
1$ (for all $B$), while $N/N_a <1$ if $Q<\pi$.}

\medskip
{\it Proof:} From (\ref{mn}), $N/N_a = 2\int_{-B}^B \s +2 \int_B^\infty
\s$ and this increases with $Q$ by (\ref{rhoincr}). Also, by
(\ref{mn}), $N/N_a = \int_{-Q}^Q\r$. When $Q=\pi$, we see from
(\ref{rho}) that $\int_{-Q}^Q\r =\int_{-\pi}^\pi (1/2\pi) =1$, so
$N/N_a =1$. To show that $N/N_a <1$ when $Q<\pi$ we use the
monotonicicity of $\s $ with respect to $B$ (Lemma 2) and $Q$ (Lemma
3) (with $\widetilde{\s}(\L) =$ the value of $\s(\L)$ for
$B=\infty, \ Q=\pi$) to conclude that  $N/N_a \leq 2\int_{-B}^B
\widetilde{\s} +2 \int_B^\infty \widetilde{\s} = \int_{-\infty}^\infty
\widetilde{\s}=1 - 2 \int_B^\infty \widetilde{\s} <1$, since
$\widetilde{\s}$ is a strictly positive function.

Finally, from (\ref{mn}) we have that $M/N_a = \int_{-B}^B \s$, and this
increases with $Q$ by  (\ref{rhoincr}). \quad {\bf QED}

\section{Solution for the half-filled band}\label{half}
In the case of a half-filled band, we have
$N=N_a, M=M'=N/2$  and, from Theorems 2 and 3, $Q=\pi, B=\infty$.
In this case the integral equations (\ref{rho}) and (\ref{sigma})
can be solved.
We use the notation $\rho_0(k)$ and $\sigma_0(\Lambda)$
for these solutions.

\medskip
Substituting (\ref{rho}) into (\ref{sigma}) where, 
as explained earlier, we use $\r_0(k)= 1/2\pi$ in the first term
on the right side of (\ref{sigma}). 
 Then the integral equation (\ref{sigma}) involves only
 $\s_0(\L)$ and can be solved by  Fourier transform.
Using equations (\ref{FT})
it is straightforward to obtain the solution for $\s$ and its 
Fourier transform $\widehat{\widehat{\s}}_0$ as 
\begin{equation}\label{sigmahat}
\widehat{\widehat{\s}}_0(\o) = \int_{-\infty}^\infty e^{i\o\L}\s_0(\L)d\L
= \frac{ J_0(\o)}{2\cosh(U\o/4)} \ , 
\end{equation}
\begin{equation}
\label{sigmasolution}
\boxed{\quad \s_0(\L) = \frac 1{2\pi} \int_0^\infty \frac {J_0(\o) \cos\,(\o \L)}
   {\cosh (\o U/4)} \, d\o \, , \quad }
\end{equation}
where 
\begin{equation}\label{bessel}
J_0(\o) = \frac{2}{\pi}\int_0^{\pi /2} \cos(\o \cos \theta )\,d\theta
= \frac{1}{\pi}\int_0^{\pi}\cos(\o\sin \theta) d\theta 
\end{equation} 
is the  zeroth order Bessel function.

Next we substitute (\ref{sigmasolution}) into (\ref{rho}) and this
leads (with (\ref{FT}))  to
\begin{equation}
\label{rhosolution}
\boxed{\quad \r_0(k) = \frac{1}{2\pi} + \frac {\cos k}{\pi} \int_0^\infty 
\frac {\cos (\o \sin k) J_0(\o) } { 1+e^{\o U/2}}\,  d\o. \quad }
\end{equation}

 The substitution of (\ref{rhosolution}) into (\ref{energy1}) finally
 yields the ground state energy, $E_0$, of the half-filled band as
\begin{equation}
\label{gstateenergy}
\boxed{ \quad  E_0\Big(\frac{N_a}{ 2}, \frac{N_a}{ 2}\Big)= 
-4 N_a \int _0^\infty \frac {J_0(\o)J_1(\o)}{\o (1+e^{\o U/2})}\,  d\o
\, .    \quad }
\end{equation}
where $J_1(\o) = \pi^{-1}\int_0^\pi \sin (\o \sin p)\sin p\,dp 
=  \o \pi^{-1}\int_0^\pi  \cos (\o \sin p)\cos^2p \, dp$ is
the Bessel function of order one.
\medskip

\noindent
{\it REMARKS:} (A.) When there is no interaction ($U=0$), $\hk$ is a
$\delta$-function; we can evaluate (\ref{sigmasolution}) and
(\ref{rhosolution}) as
\begin{align}
\s_0(\L) &= \frac{1}{2\pi \sqrt{1-\L^2}},  \quad &|\L| &\leq 1 ;
&\hskip 1truecm 
&&= 0,&\  otherwise \ ,\nonumber \\ 
\r_0(k)  &= \frac{1}{\pi},          \quad &|k| &\leq \frac{\pi}{2};
&\hskip 1truecm 
&&= 0,&\  otherwise\ . \nonumber 
\end{align}
This formula for $\r_0(k)$ agrees with what is expected for an ideal
Fermi gas.

(B.) The $U\to \infty $ limit is peculiar. From (\ref{sigmahat}) we
see that $\widehat{\widehat{\s}}_0(0) =1/2$, so $\int \s_0 =1/2$, but from
(\ref{sigmasolution}) we see that 
${\s}_0(\L) \to 0$ in this
limit, uniformly in $\L$. On the other hand $\r_0(k) \to 1/2\pi$, for
all $|k| \leq \pi$, which is what one would expect on the basis of the
fact that this `hard core' gas becomes, in effect, a one-component
ideal Fermi gas of $N=N_a$ particles.

We  now derive alternative, more revealing expressions for $\s,\r$. 

For $\s_0(\L)$ we substitute the integral representation (\ref{bessel})
for $J_0$ into (\ref{sigmasolution}) and recall the Fourier cosine 
transform (for  $\alpha >0$)
\begin{equation}\label{fourier}
\int_0^\infty \frac{\cos(\o x)}{\cosh ( \o \alpha )}
\, d\o =
\left(\frac{\pi}{2\alpha} \right)\frac{1}{\cosh(\pi x / 2 \alpha) } \ .
\end{equation}
Then, using $2 \cos a \cos b = \cos(a-b) + \cos(a+b)$ we obtain
\begin{equation}\label{sigmaeye}
\boxed{\quad \s_0(\L)=
\frac{1}{\pi U} \int_0^\pi \frac{d\theta}{\cosh[2\pi(\L+\cos \theta )/U]}
\  >0 \ .\quad }
\end{equation}

An alternate  integral representation can be derived similarly for $\r_0(k)$, 
but the derivation and the result is more complicated. 
We substitute $(1+e^x)^{-1}= \sum_{n=1}^\infty (-1)^n \exp[-nx]$, with
$x=\o U/2$, into (\ref{rhosolution}) and make use of the identity 
\cite[6.611.1]{GR} (with $\alpha = -is \pm c$ in the notation of 
\cite[6.611.1]{GR})
\begin{equation}\label{identity}
2 \int_0^\infty e^{-c\o }J_0(\o)\cos(\o s)\,d\o =
\left[(-c-is)^2+1\right]^{-1/2}+ \left[(c-is)^2+1\right]^{-1/2}\ 
\end{equation}
for $c>0$, and where the square roots $[\quad ]^{-1/2}$ in (\ref{identity})
are taken to have a positive real part.
This leads to 
\begin{eqnarray}\label{rhosum}
\r_0(k)&=& \frac{1}{2\pi}+ \frac{\cos k}{2\pi} \sum_{n=1}^\infty(-1)^{n+1}
 \Big\{ \left[(-nU/2 -i\sin k)^2+1\right]^{-1/2} \nonumber \\
&&\quad + \left[(nU/2 -i\sin k)^2+1\right]^{-1/2}\Big\} \ .
\end{eqnarray}

We can rewrite the sum of the two terms in (\ref{rhosum}) as
a single sum from $n=-\infty$ to $\infty$,
after making a correction for the $n=0$ term 
(which equals $\cos k / |\cos k|$ for $k\neq \pi/2$).  We obtain 
\begin{eqnarray}
\r_0(k)&=&\frac{1}{2\pi}\left[1+\frac{\cos k}{|\cos k|}\right] -
\frac{\cos k}{2\pi}\sum_{n=-\infty}^\infty (-1)^n\left[(nU/2-i\sin k)^2
+1 \right]^{-1/2} \nonumber \\
&=& \frac{1}{2\pi}\left[1+\frac{\cos k}{|\cos k|}\right]-
\frac{\cos k}{2\pi} \frac{1}{2\pi i} \int_C \frac{dz}{\sqrt{(zU/2-i\sin k)^2
+1}}\, \frac{\pi}{\sin(\pi z)} \ . \nonumber \\ 
&& 
\end{eqnarray}
The contour $C$ encompasses the real axis, i.e., it  runs to the right 
just below the real axis and to the left just above the real axis.

The integrand has two branch points $y_\pm$ on the imaginary axis,
where $y_\pm =(2i/U)( \sin k \pm 1)$. In order to have the correct
sign of the square root in the integrand we define the branch cuts of
the square root to extend along the imaginary axis from $y_+$ to $+
\infty$ and from $y_-$ to $-\infty$ . We then deform the upper half of
the contour $C$ into a contour that runs along both sides of the upper
branch cut and in two quarter circles of large radius down to the real
axis. In a similar fashion we deform the lower half of $C$ along the
lower cut. As the radius of the quarter circles goes to $\infty$ this
gives rise to the following expression.
\begin{equation}\label{rhoeye}
\boxed{\quad \r_0(k)= \frac{1}{2\pi}\left[1+\frac{\cos k}{|\cos k|}\right]-
\frac{\cos k}{2\pi U}\left[I_-(k) + I_+(k)\right] \ >0 \, \quad }
\end{equation}
where
\begin{equation}
I_\pm(k) = \int_{1\pm \sin k}^\infty \frac{d\alpha}{\sinh(2\pi \alpha /U)
\sqrt{(\alpha \mp \sin k)^2 -1}} \ . 
\end{equation}
By introducing the variable $\alpha= \cosh x \pm \sin k$ we finally obtain
the simple expression
\begin{equation}\label{eye}
I_\pm(k)= \int_0^\infty \frac{dx}{ \sinh\left\{ (2\pi/U)(\cosh x \pm \sin k)
\right\} } \ .
\end{equation}
As a consequence of expressions (\ref{rhoeye}) and (\ref{eye}) for
$\r$, we have the crucial bound needed as input in Lemma 3:

\medskip
\noindent {\bf Lemma 5 ($\r$ bounds)}. {\it When $B=\infty$ , $Q=\pi$, 
and $U>0$
\bea\label{rholimits}
1/2\pi<\r_0(k)<1/\pi\quad\quad &{\mathrm{for}}& \ 0\leq |k|< \pi/2
\nonumber \\ 0<\r_0(k)<1/2\pi \quad\quad &{\mathrm{for}}& \ \pi/2 <
|k|\leq \pi
\eea
Equivalently,
$f_0(x) <2t(x)$ for all $|x|\leq 1$.}

\medskip
{\it Proof:} When $ \pi/2 < |k| \leq \pi$ and $\cos k < 0$ the first
term $[\quad ]$ in (\ref{rhoeye}) is zero while the second term is
positive (since $I_\pm(k) >0$).  On the other hand, when $0\leq |k| \leq
\pi/2$, Theorem 1 shows that $\r_0(k) > 0$. Thus, we conclude that
$\r_0(k)>0$ for all $|k|\leq \pi$.

{}From (\ref{rho}) and the positivity of $\s$ we conclude that
$\r_0(k)<1/2\pi$ when $\pi/2 < |k|\leq \pi$. From the positivity of
$\r_0(k)$ when $\pi/2 \leq |k|\leq \pi$ we conclude that the integral in
(\ref{rho}) is less than $1/2\pi$ for all values of $0\leq \sin k <
1$ and, therefore, $1/2\pi<\r_0(k)<1/\pi$ for $0\leq |k|<\pi/2$.  {\bf
QED}

\section{Absence of a Mott transition}\label{mott}
A system of itinerant electrons exhibits a Mott transition if it
undergoes a conducting-insulating transition when an interaction
parameter is varied.  In the Hubbard model one inquires whether a
Mott transition occurs at some critical $U_c >0$.  Here we show that
there exists no Mott transition in the 1D Hubbard model for all $U>0$.

Our strategy is to compute the chemical potential $\mu_+$
(resp. $\mu_-$) for adding (resp. removing) one electron.  The system
is conducting if $\mu_+=\mu_-$ and insulating if $\mu_+>\mu_-$.

In the thermodynamic limit we can define $\mu$ by $\mu = dE(N)/dN$,
where $E(N)$ denotes the ground state energy with $M=M' = N/2$. As we
already remarked, this choice gives the ground state energy for all
$U$, at least in the thermodynamic limit.

The thermodynamic limit is given by the solution of the integral
equations, which we analyzed in Sect.~\ref{analysis}. In this limit
one cannot distinguish the odd and even cases (i.e., $M=M' = N/2$ if
$N$ is even or $M=M'-1=(N-1)/2$ if $N$ is odd.) and one simply has
$M/N= 1/2$ in the limit $N_a\to \infty$. In this case Theorem 2 says
that we must have $B=\infty$. Then only $Q$ is variable and Theorem 3
says that $Q$ is uniquely determined by $N$ provided $N\leq N_a$.

In the thermodynamic limit we know, by general arguments, that $E(N)$
has the form $E(N) =N_a e(N/N_a)$ and $e$ is a convex function of
$N/N_a$. It is contained in (\ref{energy1}) when $N/N_a \leq 1$. A
convex function has right and left derivatives at every point and,
therefore, $\mu_+=$ right derivative and $\mu_-=$ left derivative are
well defined. Convexity implies that $\mu_- \leq \mu_+$.

For less than a half-filled band it is clear that 
$\mu_+=\mu_-$ since $E(M,M)$ is smooth in $M=N/2$
for $N\leq N_a$. The chemical potential cannot make any jumps in this
region. 
But, for $N>N_a$  we have to use
hole-particle symmetry to calculate $E(N)$.
 The derivatives of $E(N)$, namely $\mu_+$ and $\mu_-$, can now be
different above and below the half-filling point $N =N_a$ and this
gives rise to the possibility of having an insulator. We learn from
(\ref{energyinvert}) that
\begin{equation}
\mu_++\mu_- = U,
\end{equation}
and hence $\mu_+ > \mu_-$ if $ \mu_- < U/2$.

We calculate $\mu_- $ in two ways, and arrive at the same conclusion
\begin{equation}\label{mmu}
\boxed{\quad \mu_-(U) = 2-4 \int_0^\infty \frac {J_1(\o)}
{\o [1+ \exp(\o U/2 )]}\, d\o \ . \quad}
\end{equation}
The first way is to calculate $ \mu_-$ from the integral equations by
doing perturbation theory at the half-filling point analyzed in Sect.
\ref{half}. This is a `thermodynamic' or `macroscopic' definition of $ \mu_-$
and it is given  in Sect. \ref{muint} below.
(From now on $ \mu_-$ means the value at the half-filling point.)

In Sect. \ref{mumicro} we calculate $ \mu_-$ `microscopically' by
analyzing the Bethe Ansatz directly with $N=N_a-4 $ electrons. Not
surprisingly, we find the same value of $ \mu_-$. This was the method
we originally employed to arrive at \cite[Eq. 23]{liebwu}.

Before proceeding to the derivations of (\ref{mmu}), we first show
that (\ref{mmu}) implies
$\mu_- < U/2$ for every
$U>0$. To see that $\mu_- < U/2$  we  observe that 
\bea
\mu_-(0) &=& 2-2\int_0^\infty \frac{{J_1(\o)} }{ \o} \, d\o\; = \;0 
\label{mu0} \\\mu'_-(0) &=& \frac{1}{ 2} \int_0^\infty J_1(\o) \, d\o = 
\frac{1}{ 2}\, .\label{dmu0}
\eea
Then $\mu_- < U/2$ holds if $\mu''_-(U) <0$, which we turn to next.
Here, (\ref{mu0}) is in \cite[6.561.17]{GR} and  (\ref{dmu0}) is 
in \cite[6.511.1]{GR}

Expanding the denominator in the integrand of (\ref{muminus})
and integrating term by term, we obtain
\begin{equation*}
\mu_-(U) = 2-4 \sum_1^\infty (-1)^n \Big[ \sqrt {1+\frac{{n^2U^2}}{
4}}-\frac{{nU}}{ 2} \Big]\nonumber
\end{equation*}
using which one  obtains
\bea
\mu''_-(U) &=&2 \sum_{-\infty}^\infty (-1)^n  \frac{{n^2}}{ 
{\big(1+\frac{{n^2U^2}}{ 4}\big)^{3/2}}} \nonumber \\
&=& \frac 2 {2\pi i} \int_C \frac {z^2}{\big(1+\frac{{U^2z^2}}{ 4}\big)^{3/2}}
 \cdot\frac \pi {\sin \pi z} \, dz \label{contourint}
\eea
where we have again replaced the summation by a contour integral with the
contour $C$ encompassing the real axis.  The integrand in
(\ref{contourint}) is analytic except at the poles on the real axis
and along two branch cuts on the imaginary axis.  This allows us to
deform the path to coincide the imaginary axis, thereby picking up
contributions from the cuts.  This yields
\begin{equation}
\mu''_-(U)= - \frac{{32}}{ {U^3}} \int_1^\infty \frac{{y^2}} 
{\big(y^2-1\big)^{3/2}}
\cdot \frac {1}{ {\sinh \big(\frac{{2\pi y}}{ U} \big)}} \, 
dy <\, 0\,, \hskip .5cm {\rm
for\>\>all\>\>} U>0 \, .
\end{equation}
Thus, we have established $\mu_+(U)> \mu_-(U)$, and hence the 1D
Hubbard model is insulating for all \, $U>0$.  {\it There is no
conducting-insulating transition in the ground state of the 1D Hubbard
model} (except at $U=0$).  

\subsection{Chemical potential from the integral equations}\label{muint}
As we noted, we take $B=\infty $ and $Q<\pi$. In fact we take $Q =\pi
-a$ with $a$ small. (In the notation of Sect. \ref{analysis}, $a=\sin
Q$, but to leading order in $a$, $\sin Q = \pi -Q$ and we need not
distinguish the two numbers.)  Our goal is to calculate $\delta E$, the
change in $E$ using (\ref{energy1}) and $\delta N$, the change in $N$
using (\ref{density}); $\mu_-$ is the quotient of the two numbers.

As before, we use the notation $\r(k)$ for the density at $Q =\pi -a$ and
$\r_0(k)$ for the density at $Q=\pi$, as given in (\ref{rhosolution}),
(\ref{rhoeye}). 

We start with $N$. As explained earlier, $\r-1/2\pi$ is odd around
$\pi/2$ so, from (\ref{density}),
\begin{eqnarray}\label{en}
\frac{N}{N_a}&=& \int_{-Q}^Q \r =2  \int_{0}^Q \r =2\int_{0}^a\r +2
\int_{a}^{\pi-a} \frac{1}{2\pi} \nonumber \\
&=& 2\int_{0}^a\r + \frac{1}{\pi}(\pi-2a) \approx 1+2a\left(\r_0(0)
-\frac{1}{\pi}\right)\ .
\end{eqnarray}
In the last expression we used the fact (and will use it again) that
$\r$ is continuous in $k$ and $a$ (as we see from (\ref{fiterate2}));
therefore, we can replace $\int_{0}^a\r$ by $a \r_0(0)$ to leading order
in $a$. We learn from (\ref{en}) that $\delta N/N_a = 2a \left(\r_0(0)
-1/{\pi}\right) <0$. 

The calculation of $\delta E$ is harder. From  (\ref{energy1}) 
\begin{eqnarray}\label{deltae}
\frac{E}{N_a}&=& -4\int_0^Q\r\cos k = -4\int_0^a\r\cos k -4\int_a^{\pi-a}
\r\cos k \nonumber \\
&\approx& -4a\r_0(0) -4 \int_a^{\pi-a}(\r -\frac{1}{2\pi}) \cos k -
\frac{2}{\pi} \int_a^{\pi-a} \cos k \nonumber \\
&=&  -4a\r_0(0) -8 \int_a^{\pi/2}(\r -\frac{1}{2\pi}) \cos k \nonumber \\
&=& -4a\r_0(0)+\frac{8}{2\pi} (1-\sin a) -8 \int_0^{\pi/2}\r\cos k +8
\int_0^a \r \cos k \nonumber\\
&\approx& +4a \r_0(0)-\frac{4a}{\pi} -8 \int_0^{\pi/2} \delta \r \cos k
+\frac{4}{\pi} -8\int_0^{\pi/2} \r_0 \cos k \ ,
\end{eqnarray}
where $\delta \r \equiv \r - \r_0$. The last two terms in 
(\ref{deltae}) are the energy of the half-filled band, $N=N_a$. 

Our next task is to compute $\delta \r $ to leading order in $a$.
It is more convenient to deal with the function $\delta f \equiv 
f-f_0$ and to note (from (\ref{frho})) that 
$\int_0^{\pi/2} \r(k) \cos k dk  = \int_0^1 f(x) \sqrt{1-x^2}\, dx$. 
We turn to (\ref{fiterate2}) and find, to leading order,  that
\begin{equation}\label{eff}
f\approx (1+2\hu )\,t - \hu \ha t +2\hu \ha \hu\,  t= f_0 +
\hu \ha f_0 -2\hu \ha \, t
\end{equation}
with $f_0 = (1+2\hu )\,t$. We note that $\hu = \hk^2(1+\hk^2)^{-1}$ since 
$B=\infty$ (see (\ref{U})) and has a kernel which we will call $u(x-y)$.
If $g$ is continuous near $0$ (in our case $g=f_0$ or $g=t$) then
$(\hu\ha \, g)(x) = \int_{-a}^a u(x-y)g(y)\, dy \approx 
2a u(x-0)g(0)$ to leading order in $a$. We also note
from (\ref{frho}) that $f_0(0)=\r_0(0)$.
Therefore, 
\begin{equation}\label{dint}
\int_0^{\pi/2} \delta \r \cos k \approx 
a \left[\r_0(0) -\frac{1}{\pi}\right]\int_{-1}^1 
\sqrt{1-x^2}\, u(x) \, dx\ .
\end{equation}

The integral in (\ref{dint}) is most easily evaluated using Fourier
transforms and Plancherel's theorem,
\begin{eqnarray}
\int_{-1}^1\sqrt{1-x^2}e^{i\o x}\, dx &=& 2\int_{0}^1\cos (\o x)
\sqrt{1-x^2}\, dx \nonumber \\
&=&2\int_0^{\pi/2}\cos(\o \sin \theta) \cos^2 \theta d\theta
=\frac{\pi}{\o}\, J_1(\o)
\end{eqnarray}
and from (\ref{FT})
\begin{equation}
\int_{-\infty}^\infty u(x)e^{i\o x}\, dx = \left[1+e^{|\o U/2|}\, \right]^
{-1} \ .
\end{equation}
By combining these transforms we can evaluate $\delta E$ from (\ref{deltae}).
\begin{equation}\label{finaldeltae}
\frac{\delta E}{N_a }= 2a \left[\r_0(0) -\frac{1}{\pi}\right]
\left[2-4 \int_0^\infty \frac {J_1(\o)}{\o [1+ \exp(\o U/2 )]}\, d\o
\right]
\end{equation}
By dividing (\ref{finaldeltae}) by (\ref{en}) we obtain (\ref{mmu}).

\subsection{Chemical potential from the Bethe Ansatz} \label{mumicro}
The evaluation of the chemical potentials from the Bethe Ansatz is
reminiscent of the calculation of the excitation spectrum of the 1D
delta-function Bose gas solved by one of us \cite{lieb}.  We consider
the case of a half-filled band.  To use our results in the previous
sections, which hold for $M,M'$ odd, we calculate $\mu_-$ by removing
4 electrons, 2 with spin up and 2 with spin down, from a half-filled
band.  This induces the changes
\begin{equation}
N\to N-4=N_a -4 \, ,\hskip 2cm M\to M-2=N_a/2 -2 \label{integers}
\end{equation}
The equations (\ref{finiteequ1})
and (\ref{finiteequ2}) determining the new
$k'$ and $\L'$   now read
\bea
&& N_ak'_j = 2 \pi I'_j +\sum_{\b=2}^{M-1} \t\big(2\sin\, k'_j -
2\L'_\b\big), \quad j=3,4,\cdots, N-2 \label{mottequ1} \\
&&\sum_{j=3}^{N-2} \t\big(2\sin\, k'_j - 2\L'_\a\big) = 2\pi J'_\a
-\sum_{\b=2}^{M-1}\t\big(\L'_\a-\L'_\b\big),\;
\a = 2,\cdots, M-1,\nonumber \\
\label{mottequ2} 
\eea
 Under  the changes (\ref{integers}), the values of
 $I'$ and $J'$ are the same as those of $I$ and $J$,
namely, 
 \bea
I'_j &=&  I_j \, , \hskip 1cm j=3,4,\cdots, N-2 \nonumber \\ 
J'_\a &=& J_\a \, , \hskip 1cm \a= 2,3,\cdots . M-1 \nonumber 
\eea
so they are centered around the origin with $k_{\rm
total}'=k_{\rm total}$.

The removal of four electrons  causes the
 values of $k$ and $\L$   to shift by small amounts, and we write 
\begin{equation*}
k'_j = k_j + \frac1 {N_a}\, w(k_j)\,, \hskip 1cm \L'_\a = \L_a + \frac{1}{ {N_a}}\,u(\L_a)
\end{equation*}
By taking the differences of (\ref{mottequ1}) and (\ref{finiteequ1}),
and  (\ref{mottequ2}) and (\ref{finiteequ2}), and
keeping the
leading terms, one obtains
\bea
&& \quad w(k_j)=  \frac{1}{ {N_a}}\, \sum_{\b=2}^{M-1}
\t'(2\sin k_j -2\L_\b) \big[2\cos k_j w(k_j) - 2u(\L_\b)\big], \label{mott1} \\
 && 4\t(2\L_\a)
-\frac{1}{ {N_a}}\, \sum _{j=3}^{N-2} 
\t'(2\L_\a -2\sin k_j) \big[2u(\L_\a) -2\cos k_j w(k_j)\big] \nonumber \\
   && \hskip 1cm = -\frac{1}{ {N_a}}\,\sum _{\b=2}^{M-1}\t'(\L_\a -\L_\b) 
   \big[u(\L_\a) -u(\L_\b) \big] \ . \label{mott2}
\eea

In  deriving these equations we have used facts from our analysis of the 
integral equations, namely that when $M=M'$, $-\L_1 =\L_M \approx \infty$
(i.e., $=\infty$ in the limit $N_a\to \infty$) and that when
$N=N_a$, $-k_1 = k_N \approx -k_2 =k_{N-1} \approx \pi$ as 
$N_a\to \infty$. Without using these facts there would be extra terms in
(\ref{mott1}) and (\ref{mott2}), e.g., $\theta(2\sin k_j -2\L_1)
+\theta(2\sin k_j -2\L_M)$, which is $\approx 0$ because $-\L_1 =\L_M 
\approx \infty$. 
 
\medskip
By replacing the sums by integrals and making use of 
 (\ref{contequ1}) and (\ref{contequ2}),
we are led to the coupled integral equations
\bea
&& \hskip 2cm r(k) = \int_{-\infty}^\infty K( \sin k - \L) s(\L) \,
d\L \label{mottcoupled1} \\ &&4 \t (2\L) + 2\pi s(\L) -\int_{-\pi}^\pi
K(\sin k - \L)r(k) \cos k\, dk \nonumber \\ && \hskip 2cm =
-\int_{-\infty}^\infty K^2 (\L -\L') s(\L') d\L' \label{mottcoupled2}
\eea
where
\begin{equation}
r(k) = w(k) \r_0(k)\,, \hskip 1cm s(\L) = u(\L) \s_0(\L)\, .
\end{equation}

Equations (\ref{mottcoupled1}) and (\ref{mottcoupled2}) can be solved
as follows.
Note that the third term on the left side of (\ref{mottcoupled2})
vanishes identically after substituting  (\ref{mottcoupled1}) for
$r(k)$.  Next introduce the Fourier transforms (\ref{FT}) and
\begin{equation}
\int_{-\infty}^\infty e^{i\o \L} \,\t(2\L) \, d\L = 
- \,\Big(\frac{{2\pi i}}{ \o}\Big)\, e^{-|\o| U/4},
\end{equation}
and we  obtain from (\ref{mottcoupled2})
  \begin{equation}
\label{g}
s(\L) = \frac{2}{ \pi} \int_0^\infty \frac{\sin \o \L}{ \o \cosh (\o
U/4)}
\, d\o\, \ . 
\end{equation}
Thus, from (\ref{mottcoupled1})
\begin{equation}
\label{r}
r(k)= \frac{4}{ \pi} \int_0^\infty \frac { \sin(\o \sin k) \,d\o} 
     {\o \big(1+e^{\o U/2} \big)}\, .
\end{equation}
Note that we have $r(-k)=-r(k)$ and $ s(-\L) = -s(\L)$.

\medskip
The chemical potential $\mu_-$ for a half-filled band is now computed
to be
\bea
\mu_-(U) &=& \frac{1}{ 4}\Big[E \Big(\frac{{N_a}}{ 2},
     \frac{{N_a}}{ 2}\Big) - E \Big(\frac{{N_a}}{ 2}-2,\frac{{N_a}}{
     2}-2 \Big)\Big] \nonumber \\ &=& \frac{1}{ 4}\Big[ -2
     \sum_{j=1}^{N} \cos k_j +2 \sum_{j=3}^{N-2} \cos k'_j\Big]
     \nonumber \\ &=& \frac{1}{ 4}\Big[ -2(-1-1-1-1) +
     2\sum_{j=3}^{N-2} (\cos k'_j -\cos k_j) \Big] \nonumber \\ &=& 2
     - \frac{1}{ 2} \int_{-\pi}^\pi r(k) \sin k\, dk \nonumber \\ &=&
     2-4 \int_0^\infty \frac {J_1(\o)} {\o (1+e^{\o U/2})}\, d\o \,
     , \label{muminus}
\eea
which agrees with (\ref{mmu}).

\section{Conclusion and acknowledgments}
We have presented details of the analysis of ground state properties
of the one-dimensional Hubbard model previously reported in
\cite{liebwu}.  Particularly, the analyses of the integral equations
and on the absence of a Mott transition are new which have not
heretofore appeared in print.

It is important to note  that in order to establish  that 
our solution is indeed the true ground state of the 1D Hubbard model,
it is necessary to establish the existence
of ordered real solutions to the Bethe Ansatz equations
(\ref{betheeq}) and, assuming the solution exists,
proofs of  (a.) and (b.) as listed at the end of Sect. 3.
The fulfillment of these steps remains as an open problem.
 
We are indebted to Daniel Mattis for encouraging us to investigate the
jump in the chemical potential as an indicator of the
insulator-conductor transition. We also thank Helen Au-Yang and 
Jacques Perk for helpful discussions. FYW would like to thank Dung-Hai Lee for
the hospitality at the University of California at Berkeley
and Ting-Kuo Lee for the hospitality at the National Center 
for Theoretical Sciences, Taiwan, where part of this work was carried
 out.  Work has been supported in part by NSF grants PHY-0139984,
DMR-9980440 and DMR-9971507.

\end{document}